\documentclass[acmtog]{acmart}

\AtBeginDocument{%
  }

\setcopyright{none}
\acmDOI{XXXXXXX.XXXXXXX}

\hypersetup{
    colorlinks=true,       
    urlcolor=pink,         
    linkcolor=blue,        
    citecolor=red          
}

\newif\ifdraft
\drafttrue

\ifdraft
\usepackage{xcolor}
\definecolor{newgreen}{rgb}{0.5, 0.6, 0.0}
\newcommand{\rac}[1]{{\color{teal}[\textbf{Rameen:} #1]}}
\newcommand{\opc}[1]{{\color{orange}[\textbf{Or:} #1]}}
\newcommand{\dcc}[1]{{\color{blue}[\textbf{Danny:} #1]}}
\newcommand{\kac}[1]{{\color{purple}[\textbf{Kfir:} #1]}}
\newcommand{\willi}[1]{{\color{green}[\textbf{W} #1]}}

\newcommand{\todo}[1]{{\color{blue}[TODO: #1]}}

\else
\newcommand{\rac}[1]{}
\newcommand{\opc}[1]{}
\newcommand{\dcc}[1]{}
\newcommand{\kac}[1]{}
\newcommand{\willi}[1]{}
\newcommand{\todo}[1]{}

\fi

\usepackage{amsmath}
\usepackage{booktabs}
\usepackage{multicol}
\usepackage{multirow}
\usepackage{siunitx}
\usepackage{hyperref}
\usepackage{xcolor}

\definecolor{pink}{rgb}{1.0, 0.6, 0.9} 

\hypersetup{
    colorlinks=true,
    urlcolor=pink, 
}

\begin{document}

\title{\hspace{1.7cm}  Dynamic Concepts Personalization from Single Videos}

\author{ \vspace{-0.3cm} \hspace*{0.3cm} Rameen Abdal \hspace{1.0cm} Or Patashnik \hspace{1.0cm} Ivan Skorokhodov \hspace{1.0cm} Willi Menapace \\
    \\  Aliaksandr Siarohin \hspace{0.7cm} Sergey Tulyakov \hspace{0.7cm} Daniel Cohen-Or \hspace{0.7cm} Kfir Aberman
    \\ \\  \centerline{\textit{ Snap Research}} }  

\renewcommand{\shortauthors}{Abdal et al.}

\begin{abstract}

Personalizing generative text-to-image models has seen remarkable progress, but extending this personalization to text-to-video models presents unique challenges. Unlike static concepts, personalizing text-to-video models has the potential to capture dynamic concepts -- entities defined not only by their appearance but also by their motion.
In this paper, we introduce Set-and-Sequence, a novel framework for personalizing Diffusion Transformers (DiTs)–based generative video models with dynamic concepts. Our approach imposes a spatio-temporal weight space within an architecture that does not explicitly separate spatial and temporal features. This is achieved in two key stages. First, we fine-tune Low-Rank Adaptation (LoRA) layers using an unordered set of frames from the video to learn an identity LoRA basis that represents the appearance, free from temporal interference. In the second stage, with the identity LoRAs frozen, we augment their coefficients with Motion Residuals and fine-tune them on the full video sequence, capturing motion dynamics.
Our Set-and-Sequence framework resulting in a spatio-temporal weight space effectively embeds dynamic concepts into the video model’s output domain, enabling unprecedented editability and compositionality, and setting a new benchmark for personalizing dynamic concepts.
\end{abstract}

\begin{teaserfigure}
\centering
\vspace{-0.3cm}
\centerline{\href{https://snap-research.github.io/dynamic_concepts}}{\large \textcolor{pink}{\texttt{https://snap-research.github.io/dynamic$\_$concepts}}}
\includegraphics[width=\linewidth]
			{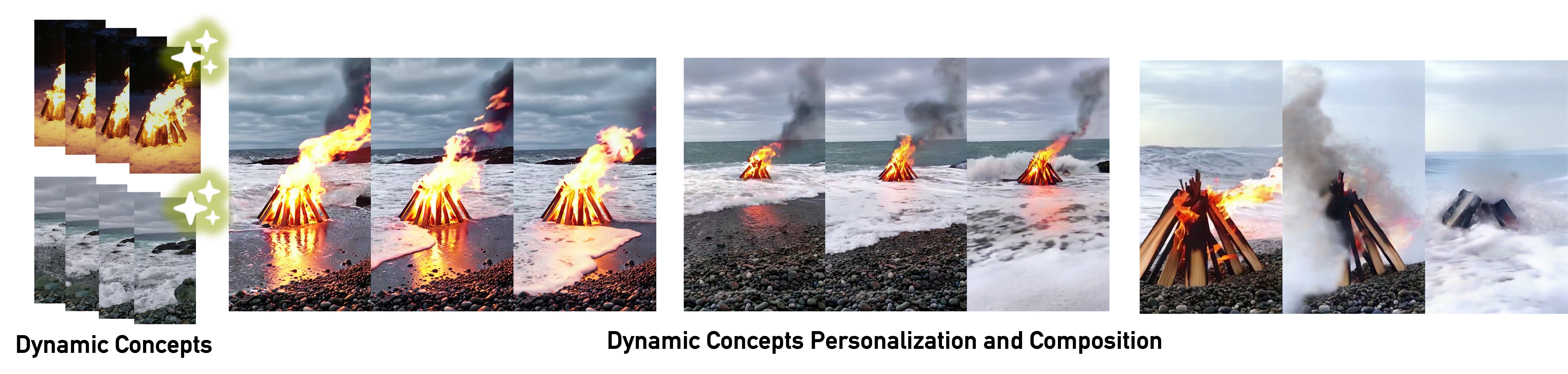}
\caption{We personalize a video model to capture \textbf{dynamic concepts} -- entities defined not only by their appearance but also by their unique motion patterns, such as the fluid motion of ocean waves or the flickering dynamics of a bonfire (left). This enables high-fidelity generation, editing, and the composition of these dynamic elements into a single video, where they interact naturally (right).}
\label{fig:teaser}
\end{teaserfigure}

\settopmatter{printacmref=false} 
\renewcommand\footnotetextcopyrightpermission[1]{} 

\maketitle


\section{Introduction}

\begin{figure*}
    \centering
    \includegraphics[width=\textwidth]{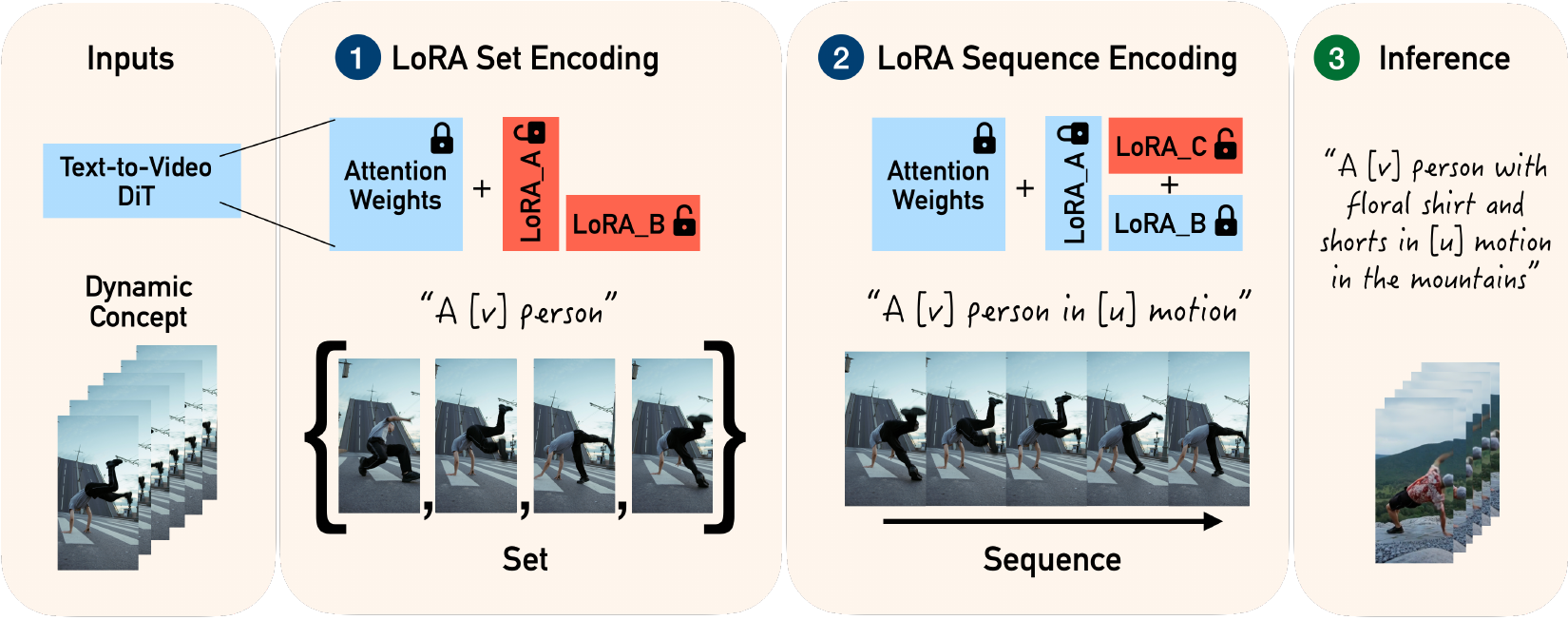} 
    \caption{{\bf Set-and-Sequence} framework operates in two stages:
(i) Identity Basis: We train \emph{LoRA Set Encoding} on a unordered set of frames extracted from the video, focusing only on the appearance of the dynamic concept to achieve high fidelity without temporal distractions.
(ii) Motion Residuals: The Basis of the Identity LoRAs is frozen and the coefficient part is augmented with coefficients of \emph{LoRA Sequence Encoding} trained on the \emph{temporal sequence} of full video clip, allowing the model to capture the motion dynamics of the concept.}
    \label{fig:pipeline}
\end{figure*}

The advent of generative models has revolutionized content creation, enabling the synthesis and manipulation of high-quality visual media with remarkable fidelity. Recent advances in text-to-image~\cite{saharia2022photorealistic,ramesh2022hierarchical,rombach2022high} and text-to-video~\cite{bar2024lumiere} models have further expanded these capabilities, opening up unprecedented opportunities for creative expression and personalization.

Personalization has been a well-established area of research in the image domain, allowing models to learn user-specific concepts that can be customized, edited, and composed into diverse contexts~\cite{ruiz2023dreambooth,gal2022image}. However, while video models have shown improvement in quality and capability, the task of personalizing these models remains an open problem. Unlike images, videos introduce an additional temporal dimension, making personalization significantly more challenging. In particular, video concepts are inherently dynamic, encompassing both appearance and motion, which must be learned and represented cohesively.

In this work, we propose a novel approach to personalizing generative text-to-video models, focusing on the idea of \emph{dynamic concepts} -- objects or subjects characterized by their entangled appearance and motion. 
Describing dynamic concepts through text is challenging, hence, embedding them into the output domain of video model and representing them as tokens,  facilitates a wide range of editing and compositional tasks.

Our work is built upon the state-of-the-art Diffusion Transformers (DiTs) architecture for video generation~\cite{menapace2024snap,sora}, which processes spatial and temporal tokens simultaneously. Unlike video generators using factorized spatial and temporal modeling (UNet-based)~\cite{blattmann2023stable} suffering from rigid artifacts, this joint spatio-temporal modeling is necessary for high-quality video generation~\cite{sora}. While DiTs achieve superior quality, they lack the innate inductive bias to disentangle spatial and temporal features. This absence of built-in separation poses a challenge for embedding dynamic concepts effectively in the model’s weight space.

Furthermore, directly fine-tuning low-rank adapters (LoRAs) on a single video often fails to capture both appearance and motion, resulting in a non-reusable representation that fails to generalize across diverse contexts or support meaningful compositionality of dynamic concepts. To address these limitations, we introduce a novel framework called \emph{Set-and-Sequence}, designed to impose a spatial-temporal structure in the weight space, enabling the representation of dynamic concepts in the weights space.

The proposed Set-and-Sequence framework operates in two stages:
(i) Identity Basis: We train LoRAs on a static \emph{unordered set} of frames extracted from the video, focusing solely on the appearance of the dynamic concept to achieve high fidelity without temporal distractions.
(ii) Motion Residuals: The Basis of the Identity LoRAs is frozen and the coefficient part is augmented with new coefficients trained on the \emph{temporal sequence} of full video clip, allowing the model to capture the motion dynamics of the concept.

This two-stage approach unlocks transformative capabilities in video generation. For the first time, we demonstrate seamless scene composition and adaptation with preserved motion and appearance. Tasks such as blending disparate video components—e.g., combining the fluid motion of ocean waves with the flickering dynamics of a bonfire—are achieved with unprecedented fidelity as shown in Fig.~\ref{fig:teaser} and in the supplementary video. Moreover, our framework enables intuitive editing of camera motion, refining expressions, and introducing localized changes, all driven by text prompts. These advancements represent a significant leap in compositionality, scalability, and adaptability, setting a new benchmark for personalized generative video models.

\section{Related Work}

\begin{figure*}[t!]
        \centering
        
        \includegraphics[width=\linewidth]{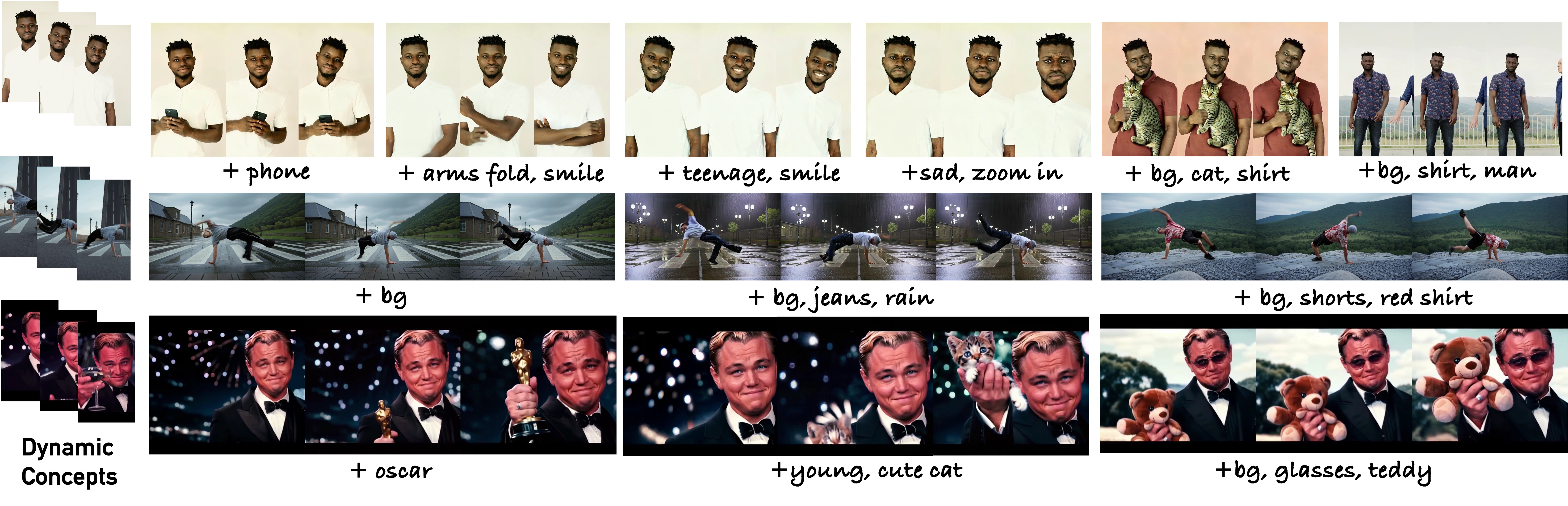}
        \caption{\textbf{Local and Global Editing.} Our \textit{Set-and-Sequence} framework enables text-driven edits of dynamic concepts while preserving both their appearance and motion. Edits can be global (e.g., background and lighting) or local (e.g., clothing and object replacement), ensuring high fidelity to the original dynamic concepts.}
        \label{fig:editing}
    \end{figure*}

\subsection{Foundational Video Models.}

Foundational video models, such as Imagen~Video~\cite{ho2022imagen}, Sora~\cite{sora}, CogVideoX~\cite{yang2024cogvideox}, Veo2~\cite{Veo2}, MovieGen~\cite{polyak2024movie} and others have made significant strides in synthesizing visually stunning and semantically aligned videos from textual descriptions. They were originally based on U-net-like~\cite{U-net} architectures~\cite{hong2022cogvideo,singer2022make,guo2023animatediff,blattmann2023stable} and were extending image generators to video synthesis by training additional temporal layers to model dynamics. However, in the pursuit of greater scalability, the community switched to transformer-based backbones with joint spatio-temporal modeling (e.g., \cite{sora,RIN,menapace2024snap}), which quickly became the dominant paradigm for large-scale video generation (e.g., \cite{polyak2024movie,sora,HunyuanVideo,yang2024cogvideox}). While these models excel at generating coherent content, they primarily rely on generic motion trajectories, limiting their ability to capture nuanced human expressions, individualized mannerisms, or complex dynamic interactions within a shared scene~\cite{sora,menapace2024snap}.
These limitations highlight the need for methods capable of personalization, dynamic scene composition, and precise editing in generative video models.
To address these challenges, we build on the video DiT (DiT version of Snap Video~\cite{menapace2024snap}) architecture and extend its capabilities with our proposed Set-and-Sequence framework, enabling the representation and compositionality of dynamic concepts with unprecedented fidelity and adaptability.

\subsection{Video Personalization and Motion Representation.}
While personalization in image generation has seen significant advancements—enabling identity preservation, stylization, and tailored manipulation~\cite{ruiz2023dreambooth,ruiz2023hyperdreambooth,gal2022image, liu2024unziploraseparatingcontentstyle,jones2024customizingtexttoimagemodelssingle}---video personalization remains relatively underexplored. In the video domain, personalization methods predominantly build upon UNet-based architectures~\cite{he2024idanimatorzeroshotidentitypreservinghuman,ma2024magicmeidentityspecificvideocustomized,zhang2024moonshotcontrollablevideogeneration,zhou2024storydiffusionconsistentselfattentionlongrange,bai2024uniedit,wu2023tuneavideooneshottuningimage}, inheriting their shortcomings.
Furthermore, approaches in this domain can be broadly categorized into three domains.
First, works like Token Flow~\cite{tokenflow2023} focus on video stylization~\cite{cai2023genren,kara2024rave,liang2023flowvid,zhang2023motioncrafter}.
Second, methods like DreamVideo~\cite{wei2023dreamvideo} and others~\cite{wu2024customttt,materzynska2024newmove,zhao2023motiondirector} emphasize extracting motion dynamics from several videos to perform motion transfer.
Third, approaches like Customize-a-Video~\cite{ren2024customize} , Fate/Zero~\cite{qi2023fatezero}, and DreamMix~\cite{molad2023dreamixvideodiffusionmodels} perform local editing on single videos by optimizing specific parts.
Although promising, these methods, such as Customize-a-Video~\cite{ren2024customize} are architecture specific and operate on the assumption that motion and appearance are disentangled, optimizing distinct LoRA~\cite{hu2021loralowrankadaptationlarge} modules or layers for each.
This rigid separation often leads to artifacts, losing fidelity and contextual realism.
Moreover, they primarily target applications like motion transfer, diverting focus from video personalization that captures the inherent entanglement of appearance and motion in dynamic concepts.
To solve this, we introduce a \textit{shared} spatio-temporal weight space that cohesively encodes dynamic concepts using a two-stage LoRA~\cite{hu2021loralowrankadaptationlarge} training.

\subsection{Scene Composition in Video Models.}
Scene composition and dynamic editing remain significant challenges in video synthesis due to the complexities of maintaining temporal coherence and contextual fidelity.
Approaches like Break-A-Scene~\cite{avrahami2023break} enable concept-level blending but are limited to static, image-like representations, relying heavily on predefined masks and cross-attention mechanisms.
In video models, scene composition often involves generating composed images using personalized text-to-image methods~\cite{qian2024omniidholisticidentityrepresentation,wang2024moamixtureofattentionsubjectcontextdisentanglement} and then applying image-to-video techniques to synthesize motion dynamics on top of the static image~\cite{chen2024videoalchemy, ren2024consisti2v,atomovideo,blattmann2023stable,dai2023animateanything,HaCohen2024LTXVideo}.
However, these models face several inherent limitations. First, they depend on powerful image composition models capable of blending multiple objects into a cohesive scene effectively ignoring the motion~\cite{blattmann2023stable,dai2023animateanything}. Second, they lack awareness of object-specific attributes, such as viewpoints, dynamic evolution of motion, and spatial relationships~\cite{blattmann2023stable,dai2023animateanything}.
Third, they fail to account for nuanced expressions and intricate motion patterns that cannot be adequately captured through text descriptions alone~\cite{dai2023animateanything,chen2024videoalchemy}.
These shortcomings render such models incompatible with the goals of video personalization and advanced compositionality.
For the first time, we demonstrate advanced compositionality by merging disparate dynamics, such as fire and water, while capturing both appearance as well as motion from single videos.
Our approach overcomes the limitations of previous methods, offering a unified framework that enables personalization of dynamic concepts.

\section{Method}

\begin{figure}[t!]
    \centering
    \includegraphics[width=\linewidth]{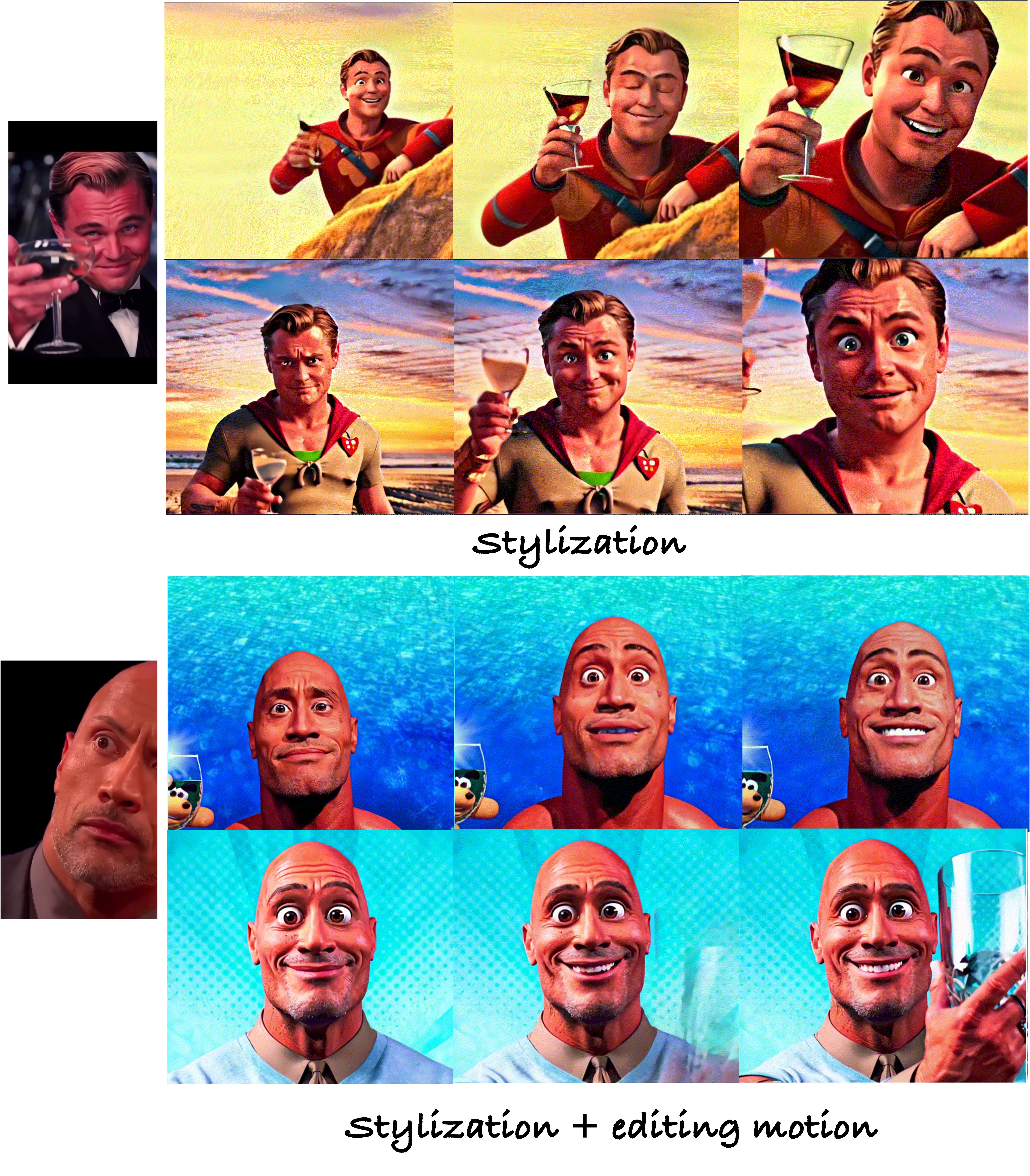}
    \caption{\textbf{Stylization.} Top: Stylization of dynamic concepts achieved by reweighting the identity basis. Bottom: Stylization and motion editing performed using prompt derived from the video in the top row. } 
    \label{fig:pixar}
\end{figure}

\begin{figure*}[t!]
    \centering
    \includegraphics[width=\linewidth]{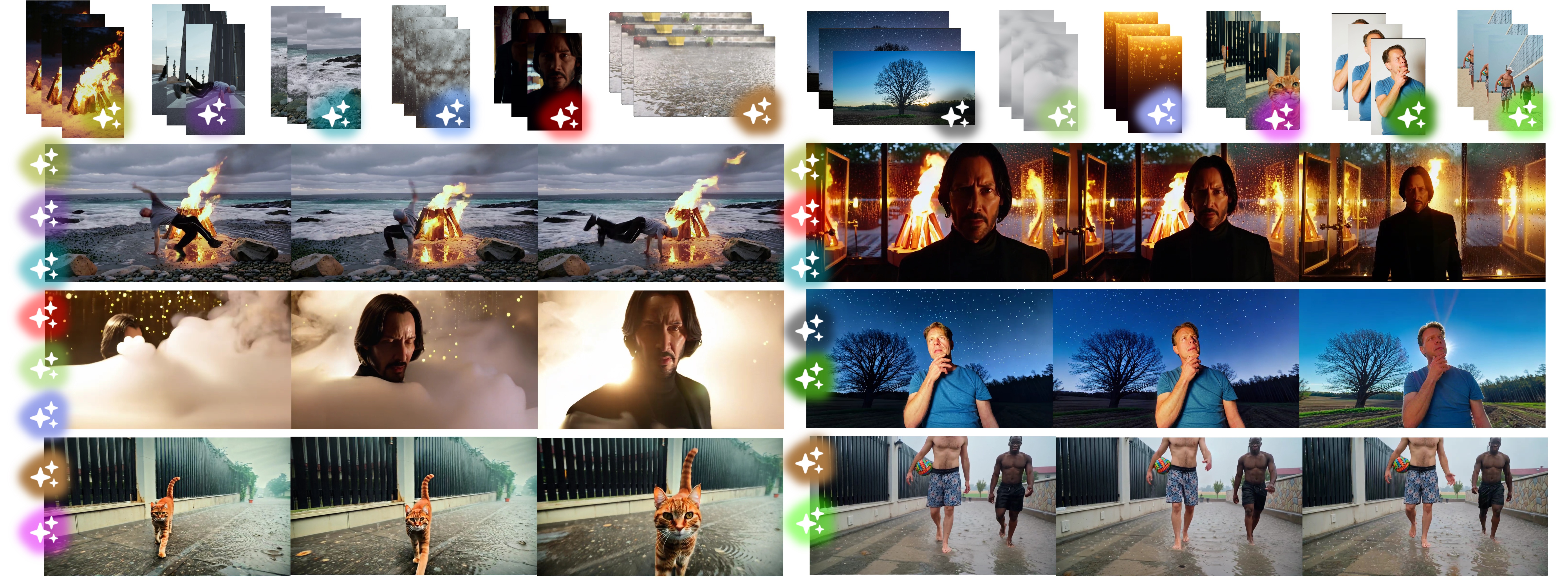}
    \caption{\textbf{Dynamic Concepts Composition.} Composition results achieved by our framework showcasing seamless integration of dynamic concepts. with each concept color-coded for clarity. For a more comprehensive demonstration, refer to the supplementary videos.}
    \label{fig:compose}
\end{figure*}

We propose \textit{Set-and-Sequence} (See Fig.~\ref{fig:pipeline}), a novel framework for personalizing text-to-video models using dynamic concepts extracted from single-video examples. Our approach learns these dynamic concepts as a decomposition of appearance and motion into a unified spatio-temporal weight space inspired by the state-of-the-art generators~\cite{menapace2024snap, sora}. We impose this weight space in DiT-based diffusion architecture~\cite{Peebles2023DiT}, an architecture that does not explicitly separate spatial and temporal features unlike UNet-based architectures~\cite{ren2024customize}, resulting in seamless compositionality, editing, and adaptation. Central to our framework is a two-stage learning technique. In the first stage, \textit{Identity Basis Learning}, we train Low-Rank Adaptation (LoRA) layers on an unordered set of video frames, extracting a static, motion-independent identity basis that captures the appearance of the concept. In the second stage, \textit{Motion Residual Encoding}, the identity basis is augmented with motion dynamics by fine-tuning coefficients on the full video sequence. We employ additional regularizations and employ text conditioning at each stage, using static prompts for appearance learning and a combination of static and dynamic prompts for encoding motion dynamics. At inference time, this enables intuitive reprompting, recomposing, and editing of content using only text descriptions, facilitating advanced personalization and dynamic scene composition.

\subsection{Preliminaries}

\paragraph{Video Diffusion with Flow Matching Loss}
Our framework builds on a video diffusion model trained with a flow matching loss~\cite{FlowStraightAndFast, NormFlowStochInterp}. This objective aligns the predicted and true velocity fields and is defined as:
\begin{equation}
\mathcal{L}_{\text{flow}} = \mathbb{E}_{\mathbf{x}, t} \left[ \left\| \mathbf{v}_\theta (\mathbf{x}_t, t) - \frac{\partial \mathbf{x}_t}{\partial t} \right\|_2^2 \right],
\end{equation}
where $\mathbf{x}_t$ represents the perturbed data at time $t$, $\mathbf{v}_\theta$ is the predicted velocity field, and $\frac{\partial \mathbf{x}_t}{\partial t}$ is the true data flow.

\paragraph{Low-Rank Adaptation (LoRA)}
LoRA fine-tunes a pretrained model by introducing low-rank updates to its weight matrices:
\begin{equation}
\mathbf{W}' = \mathbf{W} + \Delta \mathbf{W}, \quad \Delta \mathbf{W} = \mathbf{A} \mathbf{B},
\end{equation}
where $\mathbf{W}$ is the original weight matrix, $\mathbf{A} \in \mathbb{R}^{m \times r}$ and $\mathbf{B} \in \mathbb{R}^{r \times n}$ are low-rank matrices with rank $r \ll min(m,n)$. 

LoRA’s parameter efficiency and adaptability make it an ideal choice for disentangling identity and motion in video data.

\subsection{Stage 1: Identity Basis Learning}

In the first stage, we extract static identity features from an unordered set of frames as images in the input video. This stage creates a time-independent identity representation, forming the foundation for subsequent motion encoding. By decomposing and separating identity from motion, it enables the independent editing of appearance and motion during inference. The LoRA weight modification for this stage is defined as:

\begin{equation}
\mathbf{W}' = \mathbf{W} + \mathbf{A}_1 \mathbf{B}_1,
\end{equation}
where $\mathbf{A}_1 \in \mathbb{R}^{m \times r}$ and $\mathbf{B}_1 \in \mathbb{R}^{r \times n}$ represent the low-rank parameters capturing the identity. Static text tokens $\mathbf{T}_{\text{static}}$ are used to describe the subject’s appearance (e.g., as an illustration in Fig.~\ref{fig:pipeline}, “a [v] person”). In practice, for efficient editing and composition; appearance, background and expression information is also included in the static prompts to make it detailed (See supplementary materials). The [v] token is initialized with \textit{zeros}. The resulting conditional velocity field is defined as:
\begin{equation}
\mathbf{v}_\theta (\mathbf{x}_t, t; \mathbf{T}_{\text{static}}).
\end{equation}

The identity-specific flow matching loss ensures accurate reconstruction of static features. The learned parameters \(\mathbf{A}_1\) and \(\mathbf{B}_1\) are obtained by solving the following optimization problem:
\begin{equation}
(\mathbf{A}_1, \mathbf{B}_1) = \arg \min_{\mathbf{A}_1, \mathbf{B}_1} \mathbb{E}_{\mathbf{x}, t} \left[ \left\| \mathbf{v}_\theta (\mathbf{x}_t, t; \mathbf{A}_1, \mathbf{B}_1, \mathbf{T}_{\text{static}}) - \frac{\partial \mathbf{x}_t}{\partial t} \right\|_2^2 \right].
\end{equation}

This stage creates a robust, low-dimensional basis for identity representation.

\begin{figure*}[t!]
        \centering
        \includegraphics[width=\linewidth]{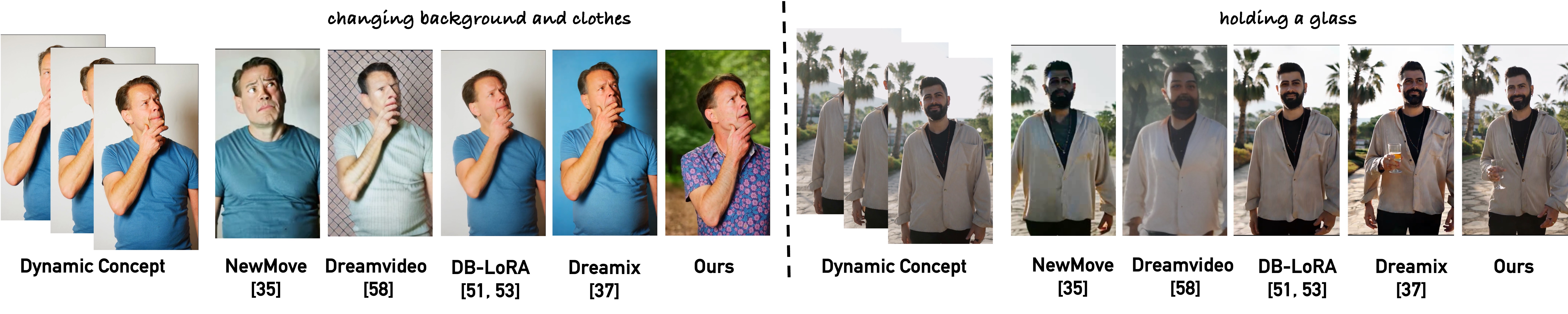}
        \caption{\textbf{Comparison with baselines.} Comparison of our method with baseline approaches (NewMove~\cite{materzynska2024newmove}, DreamVideo~\cite{wei2023dreamvideo}, DB-LoRA~\cite{simo,ruiz2023dreambooth}, and DreamMix~\cite{molad2023dreamixvideodiffusionmodels}) on two editing scenarios: changing the background and shirt, and adding a glass. Our method demonstrates superior adherence to the prompt while preserving the subject identity, outperforming the baselines.}
        \label{fig:compare}
    \end{figure*}

\subsection{Stage 2: Motion Residual Encoding}

Building upon the static identity basis established in Stage 1, the second stage introduces an additional low-rank matrix $\mathbf{B}_2$, encoding motion as a residual deformation on top of the identity. This stage captures the temporal evolution of motion dynamics, enabling independent manipulation and composition of motion during inference. The weight modification for this stage is defined as:
\begin{equation}
\mathbf{W}' = \mathbf{W} + \mathbf{A}_1 \mathbf{B}_1 + \mathbf{A}_1 \mathbf{B}_2,
\end{equation}
where $\mathbf{A}_1$ and $\mathbf{B}_1$ remain \textbf{fixed} to preserve identity, and $\mathbf{B}_2 \in \mathbb{R}^{r \times n}$ encodes motion-specific deformations. Motion encoding uses a union of static and motion-specific text tokens:
\begin{equation}
\mathbf{T}_{\text{motion}} := \mathbf{T}_{\text{static}} \cup \mathbf{T}_{\text{dynamic}},
\end{equation}
where $\mathbf{T}_{\text{dynamic}}$ describes motion attributes (e.g., as illustrated in the Fig.~\ref{fig:pipeline}, "... in [u] motion"). In practice, we also augment the dynamic part with the course action e.g., "... dancing with legs up" and camera motion e.g., "... as the camera zooms in" (See supplementary materials). The predicted velocity field is conditioned on both text components:
\begin{equation}
\mathbf{v}_\theta (\mathbf{x}_t, t; \mathbf{T}_{\text{motion}}).
\end{equation}

The flow matching loss for motion encoding ensures that the motion is reconstructed as a deformation. The learned parameters \(\mathbf{B}_2\) is obtained by solving the following optimization problem:
\begin{equation}
\mathbf{B}_2 = \arg \min_{\mathbf{B}_2} \mathbb{E}_{\mathbf{x}, t} \left[ \left\| \mathbf{v}_\theta (\mathbf{x}_t, t; \mathbf{A}_1, \mathbf{B}_1, \mathbf{B}_2, \mathbf{T}_{\text{motion}}) - \frac{\partial \mathbf{x}_t}{\partial t} \right\|_2^2 \right].
\end{equation}

This stage ensures that motion is complementary to identity, enabling robust, adaptable representations for dynamic content.

\subsection{Regularization}
\label{sec:reg}

Regularization is integral to our framework, ensuring robust training, preventing overfitting, and maintaining efficient representations of identity and motion. We employ four distinct regularization strategies to achieve these goals:

\subsubsection{Prior Preservation~\cite{ruiz2023dreambooth,roich2022pivotal}}
To ensure fidelity to the pretrained base model, we regularize both training stages to reconstruct videos generated by the base model. We sample videos with text having two parts i.e. appearance and motion, for example, in case of humans, "A man in a blue t-shirt. He is walking in a park". These prompts generalize for non-human dynamic concepts as well.

\subsubsection{High Dropout Regularization for High-Rank LoRA}
Training on a single video introduces challenges of underfitting with low-rank LoRA and overfitting with high-rank LoRA. To address these issues, we adopt a high-rank LoRA configuration combined with selective dropout applied exclusively to the $\mathbf{B}$ matrix in the LoRA update. Dropout regularization is applied to $\mathbf{B}$ LoRAs in both stages as follows:
\begin{equation}
\mathbf{B}' = \mathbf{B} \odot \mathbf{M},
\end{equation}
where $\mathbf{M}$ is a binary mask with dropout probability $p$ (e.g., $p = 0.8$). This selective dropout ensures that $\mathbf{A}_h$ remains stable, providing a consistent basis for encoding appearance and motion. By introducing sparsity in the learned coefficients, this approach mitigates overfitting while encouraging exploration of diverse parameter combinations facilitating applications like dynamic concept composition (See Fig.~\ref{fig:teaser}).

\subsubsection{Context-Aware Regularization.}

To enhance robustness and generalization, we incorporate text token masking and self-conditioning as complementary regularization techniques.

\textbf{Text Token Masking:} Random tokens in the input text are masked during training, requiring the model to infer the missing information from the remaining context. This prevents overfitting to specific token patterns and improves adaptability to diverse or incomplete prompts which we leverage for editing and re-composition.

\textbf{Self-Conditioning:} Inspired by~\cite{chen2023fit}, intermediate model outputs are reintroduced as inputs during subsequent steps, enabling iterative refinement. This feedback loop improves temporal consistency, ensuring stable identity and motion across frames.

\section{Experiment Settings}

\begin{figure}[t!]
        \centering
        \includegraphics[width=\linewidth]{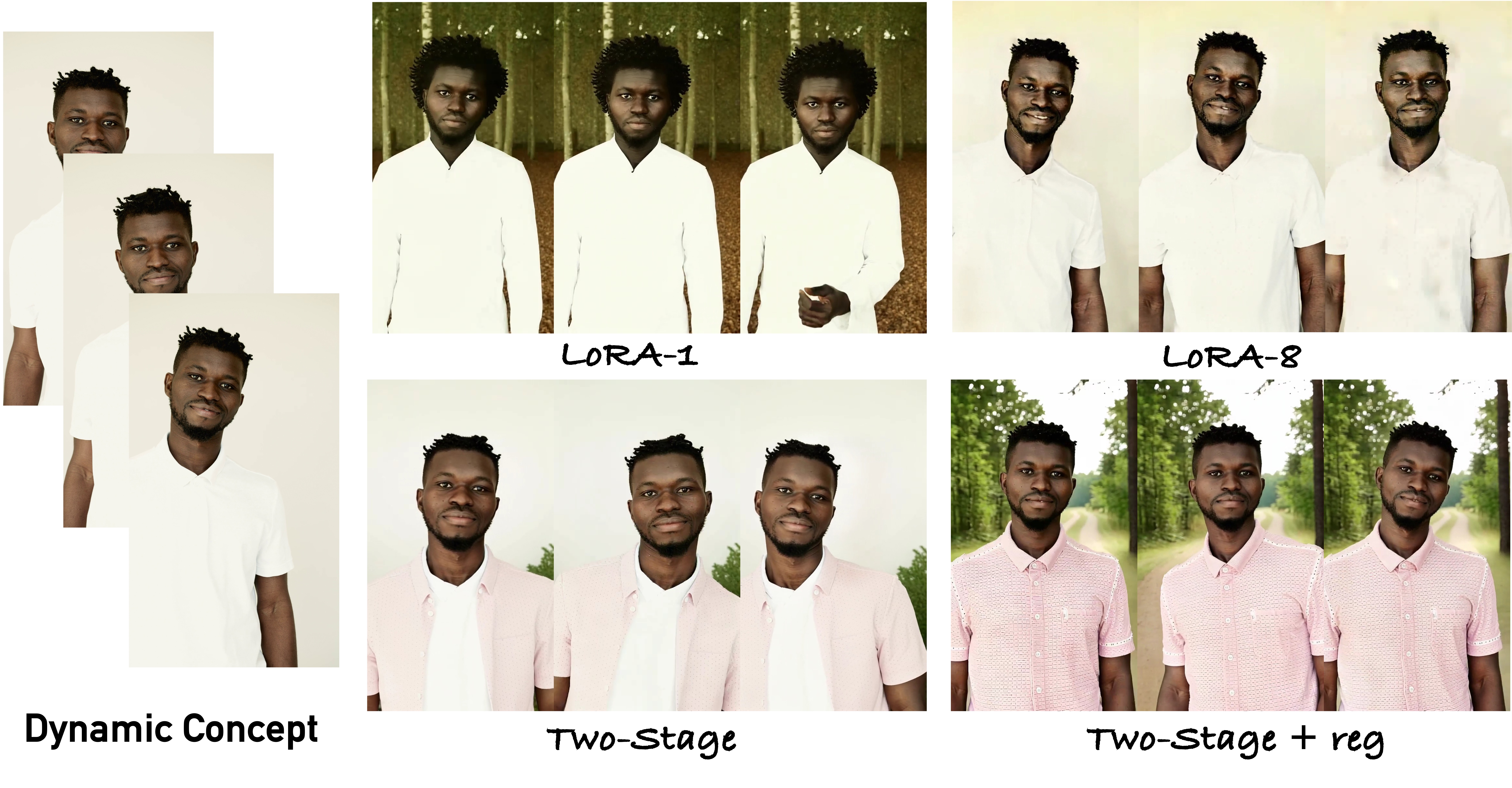}
        \caption{\textbf{Ablation.} Ablation of design choices on the editing task of adding a different shirt and background. Low-rank LoRA (LoRA-1) results in underfitting, failing to capture sufficient detail, while high-rank LoRA (LoRA-8) overfits, compromising adaptability. Our two-stage approach with added regularization achieves a balanced trade-off, preserving both fidelity and editability.}
        \label{fig:ablation}
    \end{figure}

\subsection{Evaluation Dataset} We evaluate our framework on a curated dataset of human-centric videos, as identity preservation poses a significant challenge in editing and composition. Unlike general objects, humans are highly sensitive to subtle inconsistencies in appearance, motion, and expressions, making deviations immediately perceptible. The dataset includes \textit{five} distinct identities performing actions such as dancing and walking, as well as scenario with two identities interacting in the same video. This setup tests our framework’s ability to preserve identity, maintain motion coherence, and achieve compositional fidelity compared to existing baselines. We evaluate our method using this dataset on the local and global editing tasks.

\subsection{Baselines.} We evaluate our method against several baselines, including state-of-the-art UNet-based approaches, architecture-agnostic methods, and LoRA-based variations as part of an ablation study. Among UNet-based models, we compare against DreamVideo~\cite{wei2023dreamvideo} and NewMove~\cite{materzynska2024newmove} which are state-of-the-art frameworks for video personalization. Additionally, code for methods like Customize-a-Video~\cite{ren2024customize} is unavailable, restricting direct comparison. We adapt UNet-based DreamMix~\cite{molad2023dreamixvideodiffusionmodels} (Code not available) to the DiT architecture, enabling mixed image-video training to benchmark its performance. To ensure a broader evaluation, we include architecture-agnostic methods like DreamBooth  LoRA~\cite{ruiz2023dreambooth, simo} and Textual Inversion~\cite{textual_inversion}. In our ablation study, we compare several LoRA setups to highlight the efficacy of our approach.

\subsection{Evaluation Metrics}

We employ the following metrics to quantitatively assess the quality of the generated videos:

 \textbf{Semantic Alignment.} We utilize \textit{CLIP-Text Similarity~\cite{CLIP} } \textbf{(C-T)} to measure the alignment between the generated video and the input text prompt. This metric computes the cosine similarity between the text embedding and the aggregated embeddings of all video frames, providing a global assessment of semantic consistency.

\textbf{Identity Preservation.} Maintaining identity consistency is crucial, especially in videos featuring human subjects. We utilize \textit{ArcFace Identity Similarity} (\textbf{ID})~\cite{ArcFace} to measure how well the identity of a person is preserved between the generated and reference videos. 

\textbf{Reconstruction Fidelity.} To quantify pixel-level fidelity, we compute the \textit{Mean Squared Error} (\textbf{MSE}) between corresponding frames in the generated and reference videos.

\textbf{Temporal Coherence.} Ensuring smooth transitions and motion consistency across frames is critical. For Temporal Coherence (\textbf{TC}), we compute  as CLIP image embeddings on all generated frames and report the average cosine similarity between all pairs of consecutive frames.

\section{Experiment Results}

\subsection{Quantitative Evaluation}

In order to show the effectiveness of our approach, we first ablate with the baselines and show that our two stage approach is essential for the model to integrate the dynamic concepts into the prior.  Table~\ref{tab:ablation_results} and Fig.~\ref{fig:ablation} illustrates the results of our ablation study. They are evaluated on two editing tasks i.e. \textit{"changing the shirt and background"} and \textit{"holding a glass"}.  When using a \textit{low-rank LoRA}, we observe a significant loss in identity preservation due to underfitting, as the rank is insufficient to model the complexity of dynamic concepts. Conversely, a \textit{high-rank LoRA} overfits, resulting in diminished adaptability to new prompts. In contrast, our \textit{two-stage approach} strikes a balance by using the same rank to separately train an Identity Basis and Motion Residual. This enables better adaptation to novel prompts while preserving both appearance and motion. Finally, with the added \textit{dropout regularization} discussed in Sec~\ref{sec:reg}, our framework achieves seamless integration of local edits (e.g., clothing changes) and global edits (e.g., background replacement) while maintaining motion fidelity, as shown in the supplementary videos.

To evaluate the effectiveness of our method with state-of-the-art approaches, we compare the results of various baseline methods in Fig.~\ref{fig:compare}. Each method is provided with the same editing prompt, and we assess their ability to preserve both the identity and adherence to the textual description. As shown in the figure, our framework achieves high fidelity in identity preservation and text adherence, significantly outperforming other methods. For motion preservation and coherence, we provide additional results in the supplementary videos, highlighting the seamless integration of dynamic motion with edits. Table~\ref{tab:compare} shows the quantitative analysis of the results, where our two-stage \textit{set-and-sequence} results in a better trade-off compared to other methods.
\subsection{Qualitative Results}

Our framework demonstrates significant advancements in both editing and composition of dynamic concepts, setting a new benchmark in personalized video generation. In this section, we explore two primary applications: \textit{editing} and \textit{composition}.

\subsubsection{Editing.} Our framework excels in both \textit{local} and \textit{global editing}, as demonstrated in Fig.~\ref{fig:editing}. The core objective is to capture intricate expressions and mannerisms while adapting dynamic concepts to new scenarios. For example, in Fig.~\ref{fig:editing}, our method models complex athletic movements that current video generation models fail to replicate. These results are further demonstrated in the supplementary videos, where the coherency of motion and appearance across edits is observed. Moreover, our framework supports a wide range of edits, such as changing expressions, age, and camera angles, or seamlessly integrating new objects and backgrounds into the scene. This level of control is achieved without compromising motion fidelity or identity preservation. For instance, by adjusting the weights of the LoRAs or modifying associated text prompts, our framework can produce creative outputs such as Pixar-style (See Fig.~\ref{fig:pixar}) characterizations or adapt the same dynamic concept to a completely different context. The ability to selectively adapt parts of a video, as shown in the supplementary videos, further emphasizes the flexibility of our approach.

\subsubsection{Composition.} 
\begin{figure}[t!]
        \centering
        \includegraphics[width=\linewidth]{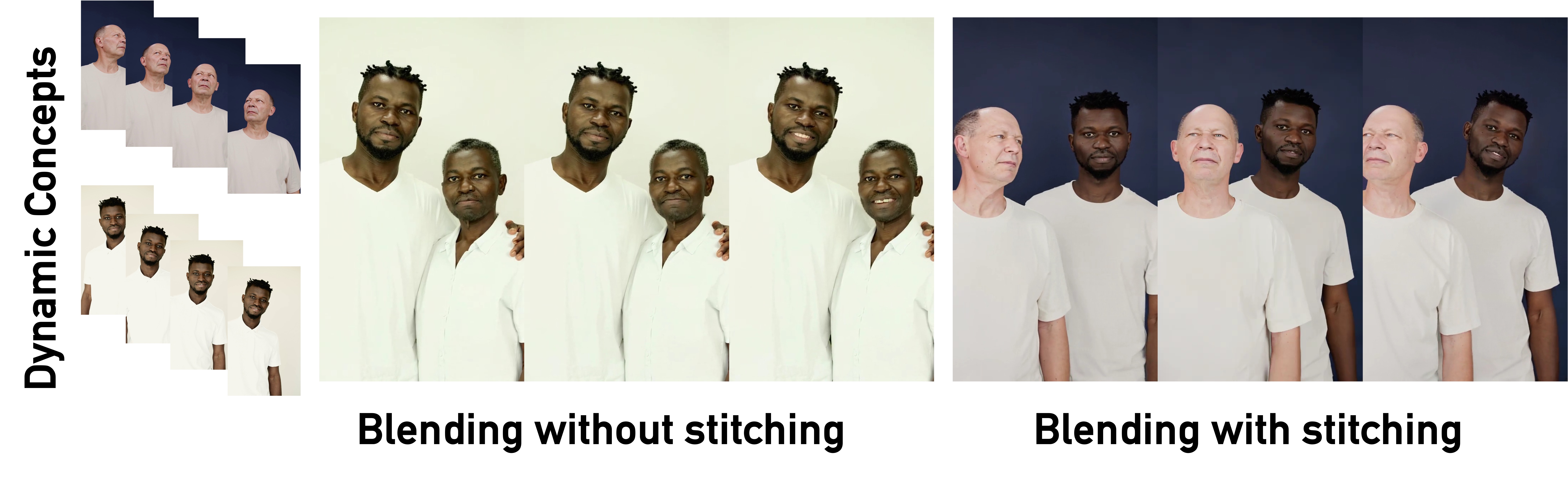}
        \caption{\textbf{Stitched Example.} To address identity leakage when generating multiple identities and motions from multiple videos (Second Column), we augment training with stitched examples by combining videos side by side to generate new compositions with preserved identities (Third Column).}
        \label{fig:stitch}
    \end{figure}

One of the most significant contributions of our framework is the ability to compose \textit{dynamic concepts} in novel and diverse settings. Leveraging the shared spatio-temporal weight space and our regularization techniques, we use single examples of multiple dynamic concepts and jointly train them using a unified identity basis and associated motion residual module. Here, after the first stage, identity basis jointly represents multiple concepts and in the second stage, the motion deformations for each dynamic concept is learned jointly. For instance, we demonstrate the composition of multiple entities, such as intricate movements, ocean waves and a bonfire, within the same scene. Fig.~\ref{fig:compose} illustrates how our method ensures coherence across these entities while maintaining their distinct characteristics. Additionally, we address challenges such as \textit{identity leakage} (see Fig.~\ref{fig:stitch}), which arises when semantically similar concepts are combined (humans in our case). To mitigate this, we employ a simple yet effective strategy of additionally training on stitched videos as a regularization. These videos are trained less frequently and although not necessary, removing backgrounds in the such stitched videos helps further in composing results involving two or more humans. More examples are provided in the supplementary videos.

\begin{table}[h]
\centering
\caption{\textbf{Ablation of Baselines.} Table evaluating Mean Square Error (MSE), Identity Preservation (ID), CLIP-T (C-T), and Temporal Coherency (TC) on the editing task. Our method demonstrates better reconstruction-edibility trade-off.}
\label{tab:ablation_results}
\begin{tabular}{lcccccc}
\toprule
\textbf{Method} & \textbf{MSE $\downarrow$} & \textbf{ID $\uparrow$} & \textbf{C-T $\uparrow$}  & \textbf{TC $\uparrow$}\\
\midrule
LoRA-1 & 0.0432  & 0.622  &0.226 & \textbf{0.9974}\\
LoRA-8  & \underline{0.0223}  & \textbf{0.703} & 0.224  & 0.9969 \\
+ Two-Stage & 0.0461  & 0.629 & \textbf{0.250} &  0.9971 \\
+ Reg       & \textbf{0.0221} & \underline{0.680}  & \underline{0.239}  & \underline{0.9972} \\

\bottomrule
\end{tabular}
\end{table}

\begin{table}[h]
\centering
\caption{\textbf{Editing Task Evaluation.} Table evaluating Mean Square Error (MSE), Identity Preservation (ID), CLIP-T (C-T), and Temporal Coherency (TC) on the editing task. Our method achieves a superior reconstruction-editability trade-off compared to the competing approaches.}
\label{tab:compare}
\begin{tabular}{lcccccc}
\toprule
\textbf{Method} & \textbf{MSE $\downarrow$} & \textbf{ID $\uparrow$} & \textbf{C-T $\uparrow$}  & \textbf{TC $\uparrow$}\\
\midrule
Tex-Inv & 0.0714  & 0.145 & 0.201  & 0.9927 \\
DB-LoRA  & \underline{0.0223}  & \textbf{0.703} & 0.224   & \underline{0.9969} \\
NewMove & 0.2223 & 0.270 & 0.204		& 0.9914 \\
DreamVideo       &0.2021  & 0.118 & 0.218	 	 & 0.9657 \\
DreamMix         & 0.0429  & 0.579  & \underline{0.226}  & 0.9965 \\
Ours              & \textbf{0.0221}   &  \underline{0.680} & \textbf{0.239}  & \textbf{0.9972}  \\
\bottomrule
\end{tabular}
\end{table}

\subsection{User Study}

To evaluate the quality of identity preservation, motion fidelity, and adherence to prompts on the editing task, we conducted a user study with 10 participants. We omit UNet based methods due to the overall lower quality (See Supplementary Videos). Participants were presented with pairs of videos generated by different methods and were asked to separately select the video that performed better in terms of identity preservation, motion fidelity, and adherence to the prompt. The results of the user study, summarized in Table~\ref{tab:user_study_results}, demonstrate that our method consistently outperforms competing approaches by achieving a better tradeoff on the editing task.

\begin{table}[h]
\centering
\caption{\textbf{User Study.} User study results comparing methods on Identity Preservation (ID), Motion Preservation (MP), Adherence to Prompt (AP), and Overall Preference of the edits (OP). Preference is computed in percentages.}
\label{tab:user_study_results}
\begin{tabular}{lcccc}
\toprule
\textbf{Method} & \textbf{IP} & \textbf{MP} & \textbf{AP} & \textbf{OP} \\
\midrule
Ours \textit{vs} DreamMix     & 87\% & 88\% & 98\% & 100\% \\
Ours \textit{vs} LoRA-1      & 99\% & 95\% & 94\% & 100\% \\
Ours \textit{vs} LoRA-8 (DB-LoRA)     & 78\% & 75\% & 98\% & 98\% \\
Ours \textit{vs} Two-Stage    & 86\% & 97\% & 76\% & 90\% \\
\bottomrule
\end{tabular}
\end{table}

\section{Limitations}

While our framework achieves state-of-the-art performance in video personalization and dynamic concept modeling, it does have limitations. The training process, which involves LoRA optimization with additional regularization, can be computationally intensive. An encoder based approach would be an ideal solution for future work. Additionally, while the method captures most motions with high fidelity, it may struggle with high-frequency or highly complex motion patterns, such as erratic or rapid movements, where temporal consistency could be further improved. These challenges present opportunities for future work to enhance efficiency, speed, and robustness in handling more dynamic scenarios.

\section{Conclusion}

We introduced a novel framework for personalized video generation that captures dynamic concepts using a two-stage \textit{Set-and-Sequence} paradigm i.e. the first stage of identity encoding and then learning coupled motion residuals on the top. By embedding these concepts into this unified spatio-temporal weight space, our method achieves high fidelity in appearance preservation, motion coherence, and text adherence, surpassing state-of-the-art baselines. The evaluations demonstrated versatility of our framework in editing and composition, while maintaining identity and motion fidelity. The ability to compose and adapt dynamic concepts in novel ways highlights the transformative potential of our approach. This work addresses long-standing challenges in video personalization and sets a new benchmark for personalized and compositional video generation.
\begin{acks}
We thank Gordon Guocheng Qian and Kuan-Chieh (Jackson) Wang for their feedback and support.
\end{acks}
\section{Architecture and Training Details}

We build our framework as a video diffusion model operating in the latent space of a video autoencoder.

The latent representation is based on a causal video autoencoder following the architecture of MAGVITv2~\cite{yu2024language}. Our autoencoder presents a high compression ratio of $8\times16\times16$ on the time and spatial dimensions, respectively, with a bottleneck dimensionality of 32 channels. Full model details are given in Table~\ref{tab:autoencoder_magvit}.

The video diffusion backbone consists in a 11.5B parameters DiT~\cite{Peebles2023DiT} detailed in Table~\ref{tab:backbone_dit}. The model is organized into 32 DiT Blocks with a hidden dimensionality of 4096 channels, each of which consists of a self-attention layer, followed by a cross-attention layer to attend to text conditioning, and a final MLP with an expansion factor of $4\times$. To further reduce the input dimensionality, a ViT-like~\cite{Dosovitskiy2021ViT} $1\times2\times2$ input patchification operation is applied, increasing the effective video compression factor to $8\times32\times32$. This allows modeling of a 121 frames $1024\times576$px video using only 9216 tokens, and enables the use of full 3D self-attention for high-quality motion modeling~\cite{polyak2024movie} without incurring a large computational penalty associated with its quadratic cost, which is further reduced by the use of a 6144 tokens attention window. Each self-attention block consists of 32 attention heads with QK-Normalization~\cite{EsserKSD3} and is augmented with 3D-RoPE~\cite{su2024roformer} embeddings, separately applied to the attention head channels in a ratio of $2:1:1$ for the temporal and spatial dimensions, respectively. Text prompts are encoded by the T5~\cite{raffel2020t5} model and combined with video tokens through the cross-attention layers. Following \cite{Peebles2023DiT}, diffusion timestep information is injected through modulation.

We perform pretraining of the model by jointly training it on a mixture of image and video data with a resolution of $512$ or $1024$ px, aspect ratios of $16:9$ and $9:16$ for videos, and $16:9$, $1:1$, and $9:16$ for image content. We adopt a progressive training strategy on the video temporal dimension, progressively extending the number of frames from 17 to 121, corresponding to 5 seconds at our fixed framerate of 24 fps. During the pretraining stage, we use the AdamW~\cite{loshchilov2018decoupled} optimizer with a fixed learning rate of $1e-4$, 10k steps warmup, a weight decay of 0.01, $\beta$ = [0.9, 0.99], $\epsilon=1e-8$ and a total of 822k training steps. Model training is accelerated through flash attention~\cite{dao2023flashattention2} with bf16 precision, and is distributed on 256 H100 GPUs using FSDP~\cite{zhao2023pytorchfsdpexperiencesscaling}.


The training details of our \textit{Set-and-Sequence} approach are summarized in Table~\ref{tab:training_stages}. The model is trained in two stages. For a single video, Stage I of the two-stage approach without regularization is trained for 150 steps, while Stage II is trained for 400 steps, requiring a total of approximately 90 minutes to converge. However, for our final method, which incorporates dropout regularization, convergence is slower. The number of training steps varies based on the complexity and number of videos. For a single video, Stage I is trained for 600 steps and Stage II for 900 steps. For multiple dynamic concepts and videos with complex motions, such as athletic dance sequences, Stage II requires extended training of 2k to 2.5k steps. To optimize text tokens effectively while avoiding overfitting, we use a lower learning rate of $1e-5$. We observe that our method is able to generalize without optimizing for these special tokens as well. Our training uses the AdamW optimizer~\cite{loshchilov2018decoupled} with a constant learning rate of $1e-4$. To ensure stable training, we set $\beta = [0.9, 0.99]$, apply a weight decay of 0.01, and use gradient clipping with a value of 0.05. Additionally, text prompt tokens are randomly dropped with a probability of 0.1. All experiments are conducted on NVIDIA A100 GPUs with 80GB of memory, using a batch size of 8. We use a  \textit{cfg} value of 8 to generate the results.



\begin{table}[h!]
\centering
\begin{tabular}{@{}ll@{}}
\toprule
\textbf{Autoencoder}      & \textbf{MAGVIT}                        \\ \midrule
Base channels             & 16                                     \\
Channel multiplier        & [1, 4, 16, 32, 64]                     \\
Encoder blocks count      & [1, 1, 2, 8, 8]                        \\
Decoder blocks count      & [4, 4, 4, 4, 4]                        \\
Stride of frame           & [1, 2, 2, 2, 1]                        \\
Stride of h and w         & [2, 2, 2, 2, 1]                        \\
Padding mode              & replicate                              \\
Compression rate          & \(8 \times 16 \times 16\)              \\
Bottleneck channels       & 32                                     \\
Use KL divergence         & \checkmark                             \\
Use adaptive norm         & \checkmark (decoder only)              \\ \bottomrule
\end{tabular}
\caption{Autoencoder and MAGVIT specifications.}
\label{tab:autoencoder_magvit}
\end{table}

\begin{table}[h!]
\centering
\begin{tabular}{@{}ll@{}}
\toprule
\textbf{Backbone}               & \textbf{DiT}                     \\ \midrule
Input channels                  & 32                               \\
Patch size                      & \(1 \times 2 \times 2\)          \\
Latent token channels           & 4096                             \\
Positional embeddings           & 3D-RoPE                          \\
DiT blocks count                & 32                               \\
Attention heads count           & 32                              \\
Window size                     & 6144 (center)                    \\
Normalization                   & Layer normalization              \\
Use flash attention             & \checkmark                       \\
Use QK-normalization            & \checkmark                       \\
Use self conditioning           & \checkmark                       \\
Self conditioning prob.         & 0.9                              \\
Context channels                & 1024                             \\ \bottomrule
\end{tabular}
\caption{Backbone and DiT specifications.}
\label{tab:backbone_dit}
\end{table}

\begin{table}[t]
\centering
\begin{tabular}{@{}lcc@{}}
\toprule
Optimizer                   & \multicolumn{2}{c}{AdamW}              \\
Learning rate               & \multicolumn{2}{c}{$1 \times 10^{-4}$} \\
LR scheduler                & \multicolumn{2}{c}{constant}           \\
Beta                        & \multicolumn{2}{c}{[0.9, 0.99]}        \\
Weight decay                & \multicolumn{2}{c}{0.01}               \\
Gradient clipping           & \multicolumn{2}{c}{0.05}               \\
Dropout (Stage I)           & \multicolumn{2}{c}{0.8}                \\
Dropout (Stage II)           & \multicolumn{2}{c}{0.5}                \\\bottomrule
\end{tabular}
\caption{Training stages and optimization settings.}
\label{tab:training_stages}
\end{table}

\section{Prompts} Providing detailed prompts at the initialization stage is crucial for achieving high-quality editing and composition. Prompts describe not only the appearance but also their environment and dynamic behavior in detail. This allows the model to align the text description with the corresponding video frames effectively, leading to better identity and motion preservation during editing and composition.

For example, in the dancer case shown in Figure 2 of the main paper, the prompt explicitly defines the appearance, surroundings, and action: "A [v] man in black track pants, gray shirt, and cap near a road on a bridge with hands down and legs up. The man is performing [u] dancing motion with hands and legs." Here, the description of the attire (black track pants, gray shirt, and cap) ensures the model captures the subject's visual identity, while the mention of the environment (a road on a bridge) provides spatial context. The prompt also includes details about the motion ("dancing motion with hands and legs"), guiding the model to encode temporal dynamics accurately.


\bibliographystyle{ACM-Reference-Format}
\bibliography{sample-base}

\end{document}